# 1932 Efforts of Persuasion on the Part of Einstein to Renounce the Cosmological Constant

Galina Weinstein

December 5, 2015

*1932 efforts of persuasion on the part of Einstein to renounce the cosmological constant after he had been stuck to it for more than a decade and Einstein's possible influence on Robertson.*

Between 1931 and 1932, Albert Einstein and Willem de Sitter found a non-static solution of the field equations of the general theory of relativity. In 1931 Einstein wrote the following line element (Einstein 1931, 236):

(1) $ds^2 = P(t)^2 d\sigma^2 - c^2 t^2$.

$P$ (scale factor) being a function of $t$ alone and $d\sigma^2$:

(1a) $d\sigma^2 = dx_1^2 + dx_2^2 + dx_3^2$.

being the line-element of a flat three-dimensional space.

In 1932 Einstein and de Sitter chose the solution with a cosmological constant $\lambda = 0$ and without introducing a spatial curvature of the universe. The Einstein-de Sitter line element was investigated by Aleksandr Friedmann in 1922 and 1924 and independently by Georges Lemaître in 1927. Hence, if the field equations of general relativity were given, the Friedman-Lemaître cosmological models resulted; the simplest of these was the Einstein-de Sitter model (Einstein 1931; Einstein and de Sitter, 1932).

Einstein and de Sitter derived the coefficient of expansion:

(2) $h^2 = \dfrac{1}{r}\left(\dfrac{dr}{cdt}\right)^2 = \dfrac{1}{3}\kappa\rho$,

which depends on the measured redshifts.

Several months later, Einstein wrote a manuscript (AE 1-115). He took as his starting point the unmodified field equations of 1915:

(3) $R_{\mu\nu} - \dfrac{1}{2} g_{\mu\nu} R = -\kappa T_{\mu\nu}$.

The left-hand side is the Einstein tensor: $R_{\mu\nu} - \frac{1}{2} g_{\mu\nu} R$.

The spatial geometry of a static spherical world is a three-dimensional sphere of radius *r* embedded in a four-dimensional Euclidean space with positive constant curvature. Hence, a static spherical world is described by the following line element:

(4) $ds^2 = d\sigma^2 - c^2 dt^2$.

(4a) $d\sigma^2 = \dfrac{dx_1^2 + dx_2^2 + dx_3^2}{\left(1 + \dfrac{kr^2}{4}\right)^2}$.

From (3) and (4) we arrive at two contradictory equations/contradictory conditions for the (constant) world radius *r* and density of matter $\rho$:

(5) $\dfrac{1}{r^2} = 0, \quad \dfrac{c^2}{r^2} = \dfrac{1}{3} c^2 \kappa \rho$.

Thus, $\rho = 0$ and equations (3) do not permit a static world with constant density. Hence, equations (3) could not be satisfied by a static spherical world of constant radius *r*, which does not change over time and in which matter has a uniform density $\rho$ (and with introducing a spatial curvature of the universe).

Adding the cosmological term $\lambda g_{\mu\nu}$ to the Einstein tensor:

(6) $\left( R_{\mu\nu} - \dfrac{1}{2} g_{\mu\nu} R \right) - \lambda g_{\mu\nu} = -\kappa T_{\mu\nu}$,

solves the problem. One then obtains:

(7) $\dfrac{1}{r^2} = \lambda, \quad \dfrac{c^2}{r^2} = \dfrac{c^2}{3} (\kappa \rho - \lambda)$.

Hence, in the case of a static spherical world the cosmological constant is a necessity.

In 1930 following Lemaître, Arthur Stanley Eddington had treated the expanding universe as having started as an Einstein static universe. The universe expanded until it asymptotically approached the de Sitter empty universe. In this model the cosmological constant played a crucial role.

In 1931 Lemaître suggested a model with point-source-creation (a "primeval atom") in which the cosmological constant again played an important role. The model had a singularity at time zero followed by rapid expansion, this being decelerated by self-gravitation leading to near-stagnation in the vicinity of the Einstein static universe, if the value of the cosmological constant $\lambda$ was suitably chosen, until the onset of accelerated expansion under cosmic repulsion.

However, in the same year shortly before Lemaître suggested his primeval atom model, Einstein found the line element (1) and dropped the cosmological constant in the dynamical case. Later in 1949, he commented that, Lemaître's arguments did not appear to him to be sufficiently convincing in view of the present state of his knowledge. Einstein held that Lemaître was founding too much upon the field equations with the cosmological constant (Weinstein 2015, 341-342, 360-361, 364-365, 371; Einstein 1949, 684-685).

In Einstein's view, the significance of his 1931 paper and the Einstein-de Sitter 1932 paper was not so much that they described new cosmological models, but that they demonstrated that the cosmological constant was unnecessary. This was an "incomparable relief," as he wrote to Richard Tolman in the summer of 1931 (Einstein to Tolman, June 27, 1931, Kragh 2013, 7). It seems that the significance of Einstein's 1932 (AE 1-115) manuscript, "On the So-called Cosmological Problem", which appeared in 1933 (in French) as "On the Cosmological Structure of Space" (Einstein 1933), was also not so much that it presented a detailed exposition of the Einstein de Sitter model, but that it demonstrated that the cosmological constant was superfluous.[1]

Einstein showed in the 1932 (AE 1-115) manuscript that if the world radius $r$ is variable over time, then the line element (4) becomes:

(8) $ds^2 = P(t)^2 d\sigma^2 - c^2 dt^2$,

where $d\sigma^2$ is given by (4a), the scale factor is $P(t) = \left(\frac{r}{r_0}\right)$ and $r_0$ is the world radius at a certain moment of time.

This is the Friedmann-Lemaître-Robertson-Walker (FLRW) metric in Cartesian coordinates.

We define $k = 1/r_0^2$ and the curvature radius is $r$.

$d\sigma^2$ defines any three-space of constant curvature $k$ (which may, without loss of generality, be restricted to the values $k = -1, 0, +1$).

In 1922 Friedmann showed that from equations (6), one obtains two equations, which are not contradictory like equations (5) (Friedmann 1922, 381):

(9) $\left(\frac{\frac{dr}{dt}}{r}\right)^2 + \frac{2\left(\frac{d^2r}{dt^2}\right)}{r} + \frac{c^2}{r^2} = \lambda, \quad \left(\frac{\frac{dr}{dt}}{r}\right)^2 + \frac{c^2}{r^2} = \frac{1}{3}(\kappa c^2 \rho + \lambda)$.

If $\lambda = 0$, we obtain (2).

---

[1] According to (O'Raifeartaigh et al 2015, 6-7), the Einstein-de Sitter paper of 1932 (Einstein and de Sitter, 1932) was a rather slight work; the 1932 (AE 1-115) manuscript contains a detailed exposition of the Einstein de Sitter model.

In the 1931 paper, Einstein showed that from equations (6), one obtains (9), and if $\lambda = 0$ the second equation leads to (2) (Einstein 1931, 236).

In the 1932 (AE 1-115) manuscript, Einstein demonstrated that without introducing a spatial curvature of the universe, equation (8) leads to the Einstein-de Sitter line element (1). From (1) in accordance with the unmodified field equations (3), one obtains two equations, similar in form to equations (9) and (5), but not contradictory like equations (5) because $r/r_0$ is changing with time:

$$(10) \quad \left(\frac{d\,r/r_0}{dt}\right)^2 + 2\,r/r_0\left(\frac{d^2\,r/r_0}{dt^2}\right) = 0, \quad \left(\frac{\frac{d\,r/r_0}{dt}}{r}\right)^2 = \frac{1}{3}\kappa c^2 \rho.$$

If we choose the above second equation, we obtain equation (2).

Hence, in the Einstein-de Sitter model, *one starts from the 1915 unmodified field equations and proceeds without the cosmological constant*. In 1931, Einstein had been inspired by Friedmann's equations (9) *to choose $\lambda = 0$* and derive equation (2) (Einstein 1931, 236). In the 1932 (AE 1-115) manuscript the cosmological constant was not included in Einstein's dynamical derivation.

Later in 1949, Einstein explained to Lemaître that after Hubble's discovery of the expansion of the universe, and since Friedmann's discovery that the field equations with no cosmological constant involve the possibility of the existence of an average (positive) density of matter in an expanding universe, the introduction of such a constant was unjustified (Einstein 1949, 684-685). Einstein's philosophical standpoint was to choose the logically simplest field equations in light of experimental discoveries (the existence of constant density of matter in an expanding universe). The field equations with no cosmological constant (3) allowed this possibility and the Einstein-de Sitter model (1) was the simplest solution, which Einstein therefore chose (Weinstein 2015, 365).

In the 1931 paper and in the 1932 (AE 1-115) manuscript, Einstein wrote the (FLRW) metric, equations (8) and (4a). In a footnote Einstein noted that a very thorough examination of the general problem [equation (8)] and its various special cases [e.g. equation (4)] has been carried out by Tolman (AE 1-115, 10; Tolman 1929).[2] Indeed, Tolman derived an equation resembling equation (8) in spherical coordinates $r$, $\theta$, $\varphi$ (Tolman 1929, 301). In 1929, however, Harvey Percy Robertson wrote the (FLRW) metric (8) in spherical coordinates $r$, $\theta$, $\varphi$, where (Robertson 1929, 826):

---

[2] See translation into English in (O'Raifeartaigh et al 2015, 46). Einstein met Tolaman in January 1931 in Pasadena and Tolaman understood German very well. Einstein wrote in his travel diary to California that in 1929 and 1930 Tolman had published five papers with different suggestions on the cosmological problem (see Tolman 1929; Nussbaumer 2014, 9).

(4b) $d\sigma^2 = \dfrac{dr^2}{1-kr^2} + r^2(d\theta^2 + \sin^2\theta\, d\varphi^2)$.

Robertson wrote in a footnote (Robertson 1929, 827-828):

> "A. Friedman, *Z. Physik*, 10, 377 (1922); 21, 326 (1924) and, more recently, R. C. Tolman, these PROCEEDINGS, 15, 297 (1929) have also attacked the problem of deriving the most general line element suitable for relativistic cosmology. But Friedman's reduction to a normal form by means of his 'assumptions of the second class' is unsatisfactory and, it seems to the present author, is not possible on his assumptions alone. Tolman, on the other hand, has restricted the form of the line element a priori and without taking full advantage of the isotropy which he mentions, and cannot deal with non-stationary possibilities (cf. loc. cit., p. 304). Both introduce untenable assumptions on the matter-energy tensor…. and require that Einstein's field equations be satisfied instead of making full use of the intrinsic uniformity of such a space, as we do here. The non-stationary solution found by Friedman is contained in […the (FLRW) metric] as the case in which the proper pressure *p* vanishes".

After 1933, Robertson had a long period of contact with Einstein at Princeton. Having an office in the Institute for Advanced Study in Princeton next to Einstein's office, Robertson discussed cosmological matters with Einstein. It is reasonable to assume that Einstein and Robertson also discussed the (FLRW) metric.

In 1939 Einstein noted that his investigation on the Schwarzschild singularity arose out of discussions he conducted with Robertson and others (Einstein 1939, 936). At the end of the paper, "On Gravitational Waves", Einstein added a note that the second part of the paper on cylindrical waves had been considerably altered by him and he said he wished to thank his colleague, Robertson, for his friendly assistance in the clarification of an original error (Einstein and Rosen 1937, 54).

In 1935 Robertson published a major paper under the title "Kinematics and World-Structure". It is reasonable to assume that Robertson's derivation of an Einstein-de Sitter metric in this paper was influenced by his possible conversations with Einstein. Robertson demonstrated that for $k = 0$, equation (8) leads to equation (1) where $d\sigma^2$ is given by (1a) (Robertson 1935, 285-286).